\documentclass[sigplan, screen]{acmart}

\copyrightyear{2025}
\acmYear{2025}
\setcopyright{cc}
\setcctype{by}
\acmConference[ASPLOS '25]{Proceedings of the 30th ACM International Conference on Architectural Support for Programming Languages and Operating Systems, Volume 2}{March 30-April 3, 2025}{Rotterdam, Netherlands}
\acmBooktitle{Proceedings of the 30th ACM International Conference on Architectural Support for Programming Languages and Operating Systems, Volume 2 (ASPLOS '25), March 30-April 3, 2025, Rotterdam, Netherlands}\acmDOI{10.1145/3676641.3715999}
\acmISBN{979-8-4007-1079-7/2025/03}

\usepackage{eulervm}
\usepackage[T1]{fontenc}

\usepackage{anyfontsize}

\usepackage{changepage}
\usepackage{tikz}
\usetikzlibrary{matrix}
\usetikzlibrary{arrows.meta, positioning, shapes.geometric, decorations.markings, calc}
\usetikzlibrary{intersections}
\usepackage{isabelle}
\usepackage{isabellesym}

\usepackage{listings}
\usepackage{xcolor}
\usepackage{amsmath}    
\usepackage{amsthm}     
\usepackage{mathtools}  
\usepackage{mathpartir}
\usepackage{stmaryrd}
\usepackage{balance}

\usepackage{soul}
\usepackage{fontawesome}

\makeatletter
\@namedef{T1/zi4/m/it}{<->ssub*zi4/m/n}
\@namedef{T1/zi4/b/it}{<->ssub*zi4/b/n}
\makeatother

\usepackage{etoolbox}
\tikzset{
    redtrans/.style={
        ->,
        draw=red,
        text=red
    },
    bluetrans/.style={
        ->,
        draw=blue,
        text=blue
    },
    main node/.style={
        draw,
        align=center
    }
}

\definecolor{colorADDEDfg}{HTML}{238b45}
\definecolor{colorADDEDbg}{HTML}{cef0d8}
\definecolor{colorDELETEDfg}{HTML}{969696}

\definecolor{colorsketch}{HTML}{2541db}
\definecolor{colorsledge}{HTML}{c95faf}
\definecolor{colorsledgebg}{HTML}{f7e1f2}

\newcommand\CXLcache{\texttt{CXL.cache}}
\newcommand\CXLio{\texttt{CXL.io}}
\newcommand\CXLmem{\texttt{CXL.mem}}

\DeclareRobustCommand{\textadded}[1]{\textcolor{colorADDEDfg}{\sethlcolor{colorADDEDbg}\hl{#1}}}

\DeclareRobustCommand{\textdeleted}[1]{\textcolor{colorDELETEDfg}{\st{#1}}}

\newcommand{\inv}{\texttt{inv}}
\newcommand\eqdef{\overset{{\rm def}}{=}}
\newcommand\var[1]{\mathit{#1}}
\newcommand\field[1]{\textsf{#1}}
\newcommand\const[1]{\textsf{#1}}
\newcommand\lrecord{\llparenthesis}
\newcommand\rrecord{\rrparenthesis}

\newcommand\List[1]{#1\,\textrm{list}}

\newcommand\SWMR{\mathrm{SWMR}}

\newcommand\STATE[1]{\textsf{#1}}
\newcommand\EM{\STATE{M}}
\newcommand\SH{\STATE{S}}
\newcommand\I{\STATE{I}}

\usepackage{listings}
\usepackage{xcolor}
\newcommand{\buggyrules}{\texttt{BuggyRules.thy}}
\newcommand{\transposed}{\texttt{Transposed.thy}}
\newcommand{\coherenceproperties}{\texttt{CoherenceProperties.thy}}
\newcommand{\basicinvariants}{\texttt{BasicInvariants.thy}}
\newcommand{\typeonestate}{\texttt{Type1State}}
\newcommand{\swmrstatemachine}{\texttt{SWMR\_state\_machine}}

\newcommand\TSTATE[2]{\STATE{#1}^{\STATE{#2}}}
\newcommand\ISD{\TSTATE{IS}{D}}
\newcommand\IMD{\TSTATE{IM}{D}}
\newcommand\ISA{\TSTATE{IS}{A}}
\newcommand\IMA{\TSTATE{IM}{A}}
\newcommand\SMA{\TSTATE{SM}{A}}
\newcommand\SMD{\TSTATE{SM}{D}}
\newcommand\SMAD{\TSTATE{SM}{AD}}

\newcommand\MIA{\TSTATE{MI}{A}}

\newcommand\ISDI{\TSTATE{IS}{DI}}
\newcommand\IIA{\TSTATE{II}{A}}
\newcommand\SIA{\TSTATE{SI}{A}}
\newcommand\SIAC{\TSTATE{SI}{AC}}
\newcommand\ISAD{\TSTATE{IS}{AD}}
\newcommand\IMAD{\TSTATE{IM}{AD}}
\newcommand\SAD{\TSTATE{S}{AD}}
\newcommand\SA{\TSTATE{S}{A}}
\newcommand\SD{\TSTATE{S}{D}}
\newcommand\MAD{\TSTATE{M}{AD}}
\newcommand\MD{\TSTATE{M}{D}}
\newcommand\ID{\TSTATE{I}{D}}
\newcommand\MA{\TSTATE{M}{A}}

\newcommand\StableState{\var{StableState}}
\newcommand\DeviceTransientState{\var{DTransientState}}
\newcommand\HostTransientState{\var{HTransientState}}

\newcommand\Val{\var{Val}}
\newcommand\UTID{\var{Tid}}
\newcommand\utid{\var{txid}}

\newcommand\DTHReqType{\var{D2HReqType}}
\newcommand\RdShared{\const{RdShared}}
\newcommand\RdOwn{\const{RdOwn}}
\newcommand\CleanEvict{\const{CleanEvict}}
\newcommand\DirtyEvict{\const{DirtyEvict}}
\newcommand\CleanEvictNoData{\const{CleanEvictNoData}}

\newcommand\RdCurr{\const{RdCurr}}
\newcommand\RdAny{\const{RdAny}}
\newcommand\RdOwnNoData{\const{RdOwnNoData}}
\newcommand\ItoMWr{\const{ItoMWr}}
\newcommand\WrCur{\const{WrCur}}
\newcommand\CLFlush{\const{CLFlush}}
\newcommand\WOWrInv{\const{WOWrInv}}
\newcommand\WOWrInvF{\const{WOWrInvF}}
\newcommand\WrInv{\const{WrInv}}
\newcommand\CacheFlushed{\const{CacheFlushed}}

\newcommand\DeviceState{\var{DState}}
\newcommand\HostState{\var{HState}}
\newcommand\DevCache{\var{DCache}}
\newcommand\HostCache{\var{HCache}}
\newcommand\SystemState{\var{SystemState}}

\newcommand\DTHRespType{\var{D2HRspType}}
\newcommand\RspIHitSE{\const{RspIHitSE}}
\newcommand\RspIFwdM{\const{RspIFwdM}}
\newcommand\RspSFwdM{\const{RspSFwdM}}

\newcommand\RspIHitI{\const{RspIHitI}}
\newcommand\RspVHitV{\const{RspVHitV}}
\newcommand\RspSHitSE{\const{RspSHitSE}}
\newcommand\RspVFwdV{\const{RspVFwdV}}

\newcommand\HTDReqType{\var{H2DReqType}}
\newcommand\SnpData{\const{SnpData}}
\newcommand\SnpInv{\const{SnpInv}}
\newcommand\SnpCur{\const{SnpCur}}

\newcommand\HTDRespType{\var{H2DRspType}}
\newcommand\GO{\const{GO}}
\newcommand\GOShared{\const{GO-Shared}}
\newcommand\GOModified{\const{GO-Modified}}
\newcommand\WritePull{\const{WritePull}}
\newcommand\GOWritePull{\const{GO\_WritePull}}
\newcommand\GOWritePullDrop{\const{GO\_WritePullDrop}}
\newcommand\GOErrWritePull{\const{GOErrWritePull}}
\newcommand\ExtCmp{\const{ExtCmp}}
\newcommand\FastGOWritePull{\const{FastGOWritePull}}

\newcommand\DTHReqI{\field{D2HReq1}}
\newcommand\DTHReqII{\field{D2HReq2}}
\newcommand\DTHRspI{\field{D2HRsp1}}
\newcommand\DTHRspII{\field{D2HRsp2}}
\newcommand\DTHDataI{\field{D2HData1}}
\newcommand\DTHDataII{\field{D2HData2}}
\newcommand\HTDReqI{\field{H2DReq1}}
\newcommand\HTDReqII{\field{H2DReq2}}
\newcommand\HTDRspI{\field{H2DRsp1}}
\newcommand\HTDRspII{\field{H2DRsp2}}
\newcommand\HTDDataI{\field{H2DData1}}
\newcommand\HTDDataII{\field{H2DData2}}
\newcommand\devcacheI{\field{DCache1}}
\newcommand\devcacheII{\field{DCache2}}
\newcommand\hcache{\field{HCache}}
\newcommand\DBufferI{\field{DBuffer1}}
\newcommand\DBufferII{\field{DBuffer2}}
\newcommand\DProgI{\field{DProg1}}
\newcommand\DProgII{\field{DProg2}}
\newcommand\Counter{\field{Counter}}
\newcommand\State{\field{State}}
\newcommand\val{\field{Val}}
\newcommand\devcachei[1]{\field{DCache}_{#1}}
\newcommand\dthdatai[1]{\field{D2HData}_{#1}}
\newcommand\htddatai[1]{\field{H2DData}_{#1}}
\newcommand\DTHReq{\var{D2HReq}}
\newcommand\DTHResp{\var{D2HRsp}}
\newcommand\HTDReq{\var{H2DReq}}
\newcommand\HTDResp{\var{H2DRsp}}
\newcommand\Buffer{\var{DBuffer}}

\newcommand\Data{\var{Data}}

\newcommand{\SB}{\TSTATE{S}{B}}
\newcommand{\MB}{\TSTATE{M}{B}}
\newcommand{\IB}{\TSTATE{I}{B}}

\newcommand{\dotfield}[2]{#1.\field{#2}}
\newcommand{\supersketch}{\texttt{super\_sketch}}

\newcommand\Modified{\STATE{M}}
\newcommand\Exclusive{\STATE{E}}
\newcommand\Shared{\STATE{S}}
\newcommand\Invalid{\STATE{I}}

\usepackage{ifthen}

\newcommand{\makestep}[2]{%
  \xrightarrow[%
    {\mathclap{\scriptscriptstyle
      \ifstrempty{#2}{}{#2}
    }}%
  ]{%
    {\mathclap{\scriptscriptstyle
      \ifstrempty{#1}{}{#1}
    }}%
  }%
}

\newcommand\head{\var{head}}
\newcommand\tail{\var{tail}}

\newcommand\Load{\texttt{Load}}
\newcommand\Evict{\texttt{Evict}}
\newcommand\Instruction{\var{Instruction}}

\newcommand\Store{\texttt{Store}}
\newcommand\EBuffer{\const{EmptyBuffer}}
\usepackage[normalem]{ulem}
\usepackage{cancel}
\newcommand{\msout}[1]{\text{\sout{\ensuremath{#1}}}}

\usepackage{subfig}
\usepackage{afterpage}
\usepackage{quoting}
\usepackage[nohyperlinks, printonlyused, withpage, nolist]{acronym}
\usepackage{hyperref}
\usepackage{url}
\usepackage[hyphenbreaks]{breakurl}
\usepackage{listings}
\usepackage{wrapfig,booktabs}
\usepackage{tabularx}
\usepackage{cleveref}

\newlength{\leftbarwidth}
\newlength{\leftbarsep}
\usepackage{ragged2e,rotating}
\newcolumntype{C}{>{\hspace{0pt}%
    \Centering\arraybackslash}X}
\usepackage[skip=0.333\baselineskip]{caption}

\hypersetup{
  colorlinks = true,
  linkcolor = blue,
  citecolor = blue,
  urlcolor = blue
}

\usepackage{framed}
\definecolor{shadecolor}{HTML}{EEEEEE}

\newcommand\cxlcitesection[1]{\cite[\S#1]{cxl31}}

\colorlet{leftbarcolor}{black!10}

\renewenvironment{leftbar}{%
    \MakeFramed {\advance \hsize -\width \FrameRestore }%
}{%
    \endMakeFramed
}
\newenvironment{Quote}{\begin{leftbar}\noindent}{\end{leftbar}}

\definecolor{navy}{RGB}{0,0,128}

\begin{document}
\title{Formalising CXL Cache Coherence}

\author{Chengsong Tan}
\orcid{0009-0008-7822-8407}
\authornote{Work done while the author was at Imperial College London.}
\affiliation{%
  \institution{Kaihong}
  \city{Shenzhen}
  \country{China}
}

\author{Alastair F. Donaldson}
\orcid{0000-0002-7448-7961}             
\affiliation{%
  \institution{Imperial College London}
  \city{London}
  \country{UK}
}

\author{John Wickerson}
\orcid{0000-0001-6735-5533}
\affiliation{%
  \institution{Imperial College London}
  \city{London}
  \country{UK}
}

\renewcommand{\shortauthors}{Chengsong Tan, Alastair F. Donaldson, and John Wickerson}

\ccsdesc[500]{Computer systems organization}
\ccsdesc[500]{Computer systems organization~Architectures}
\ccsdesc[500]{Theory of computation~Logic and verification}

\ccsdesc[500]{Computer systems organization}
\ccsdesc[500]{Computer systems organization~Architectures}
\ccsdesc[500]{Theory of computation~Logic and verification}

\keywords{CXL, Cache Coherence, Proof Assistant, Heterogeneous Computing, Formal Proof}

\begin{abstract}
We report our experience formally modelling and verifying \CXLcache{}, the inter-device cache coherence protocol of the Compute Express Link standard. We have used the Isabelle proof assistant to create a formal model for \CXLcache{} based on the English prose specification.
This led to us identifying and proposing fixes to several parts of the specification that were unclear, ambiguous or inaccurate.
Nearly all our issues and proposed fixes have been confirmed and tentatively accepted by the CXL consortium for adoption, save for one which is still under discussion.
To validate the faithfulness of our model we performed scenario verification of essential restrictions such as ``Snoop-pushes-GO'', and used the Isabelle proof assistant to produce a fully mechanised proof of a coherence property of the model.
The considerable size of this proof, comprising tens of thousands of lemmas, prompted us to develop new proof automation tools, which we have made available for other Isabelle users working with similarly cumbersome proofs.
\end{abstract}

\maketitle 


\section{Introduction}

Compute Express Link (CXL)~\cite{cxl} is an emerging standard that provides cache coherence
across multiple devices connected along a PCIe bus. 
Inter-device cache coherence is a boon to computer architects because it allows multiple devices to communicate with each other while transferring a minimal amount of data between them.
CXL has the potential to be faster 
than other memory expansion methods \cite{DirectCXL} 
and save stranded memory in cloud computing clusters \cite{Pond}. 

CXL is not the first standard for inter-device cache coherence~\cite{opencapi, gen-z, ccix, infiniband, infinityfabric, omnipath, ultrapath, nvlink}, but it is the first to enjoy broad support across the computer industry, with backers including Alibaba, AMD, Arm, Broadcom, Cisco, Dell, Ericsson, Google, Hewlett Packard, Huawei, IBM, Intel, Meta, Microsoft, Nvidia, Oracle, Qualcomm, Samsung, Synopsys, Xilinx, and many others.

The CXL standard is large, complex and new, and is set to form a trusted pillar of datacenter computers for years to come.\footnote{Yole Group anticipates a CXL market size of \$15.8 billion by 2028~\cite{yole_cxl}.} As such, now is the ideal time to study the standard intensively. Does it contain inconsistencies? Is the wording unambiguous throughout? And perhaps most importantly: does it actually provide its stated guarantee of inter-device cache coherence?

We report here on our efforts to answer those questions.

\paragraph{Contribution 1: Formalising \CXLcache{}}
The part of the CXL standard that provides inter-device cache coherence is called \CXLcache{}. (The other two parts of the standard are \CXLio{}, which governs bulk data transfers, and \CXLmem{}, which relates to disaggregated memory.) Our first contribution is a formalisation of the \CXLcache{} protocol in the Isabelle proof assistant~\cite{isabelle}. Our formalisation takes the form of a state-transition system. It comprises a detailed model of the whole-system state (encompassing the state of caches in devices plus the contents of the various channels that contain messages sent between the devices and the `host'), together with dozens of transition rules that define the legal ways the state can evolve in response to CXL messages being passed around and processed. 

We explain in \Cref{sec:model} how our formalisation corresponds to the informal prose given in the official CXL standard, and which assumptions we have made in our modelling process.

As a direct result of our modelling efforts, we uncovered \hyperref[sec:fixingproblems]{five} areas where the CXL standard could be improved (\hyperref[sec:fixingproblems:inaccuracy]{one} inconsistency, \hyperref[sec:fixingproblems:redundancy]{one} redundancy, \hyperref[sec:fixingproblems:inefficiency]{one} inefficiency, and \hyperref[sec:fixingproblems:unclarity]{two} places where the intention could be clarified). We have proposed corresponding improvements to the text to the engineers who lead the drafting of the protocol.
In four cases they have confirmed that these will be incorporated into the next version of the standard, with one of our proposed fixes still under discussion.


\paragraph{Contribution 2:  Proving a cache coherence property}
Our second contribution involves putting our formalisation to work. 

First, we use it for a form of `scenario verification': we use a number of litmus tests (some derived from message-sequence charts in the CXL specification) and run them through our model to confirm that legal interactions are indeed allowed and illegal interactions are indeed forbidden. This helps gain confidence in the faithfulness of our model to the standard. 
We also use scenario verification to show that if some requirements imposed by the standard are relaxed, then coherence is not met---this establishes confidence that the CXL standard is not overly `strong'; i.e., that it does not impose requirements on CXL implementations without good reason.

Second, we prove that it satisfies the `single writer, multiple reader' (SWMR) property~\cite[p.~11]{primer}.
The SWMR property states that if one device has write access to a location, then no other device can simultaneously have read or write access to the same location.
SMWR is one of the two properties that are, together, sufficient to establish cache coherence; the other is the `data-value invariant'~\cite[p.~13]{primer}, which we leave as future work.

\newcommand\conjunctcount{796}
\newcommand\rulecount{68}
\newcommand\strengthenedrulecount{14}
\newcommand\conjunctrulecount{53,332}
\newcommand\filecount{73}

\paragraph{Contribution 3: Better automation for large proofs}
Our proof that our model of CXL satisfies the SWMR property is large. SWMR is not inductive on its own, so to complete the proof, we needed to devise a stronger invariant (one that implies SWMR), and prove that this invariant holds for all legal initial states of the system and is preserved by every transition rule.
%
%
The invariant is made up of \conjunctcount{} conjuncts, and there are \rulecount{} transition rules, hence we must prove \conjunctrulecount{} lemmas of the form given in \Cref{fig:lemma_inv_pres}.

\begin{figure}[t]
\begin{isabelle}
\isacommand{lemma} inv\_preservation\_$i$\_$j$: \\
\hspace{1em}\isacommand{fixes} $\Sigma$, $\Sigma'$ :: state \\
\hspace{1em}\isacommand{assumes} inv\_1($\Sigma$) $\land$ $\ldots$ $\land$ inv\_${\tt \conjunctcount}$($\Sigma$) \\
\hspace{1em}\isacommand{assumes} rule\_$i$($\Sigma$,$\Sigma'$) \\
\hspace{1em}\isacommand{shows} inv\_$j$($\Sigma'$)
\end{isabelle}
\caption{Lemmas of this form state that all rules preserve all conjuncts of the invariant. They assume all \conjunctcount{} conjuncts hold in state $\Sigma$, and that the $i$th rule can evolve the state to $\Sigma'$, then show that the $j$th conjunct of the invariant still holds.
}
\label{fig:lemma_inv_pres}
\Description{}
\end{figure}

Most of these obligations can be automatically discharged via a single call to Isabelle's \emph{sledgehammer}~\cite{sledgehammer}, but this process still requires manual intervention to copy the proof snippet discovered by sledgehammer into the overall theory file. The situation is worsened by the fact that this is not a one-shot effort: our invariant had to be revised many times during the proof development process, because we frequently found that an extra conjunct was needed in order to make one of the lemmas hold. Moreover, each time a conjunct was added, it was necessary to show that it is preserved by all of the transition rules, which in turn often led to the need for further conjuncts!

As such, robust proof automation is a necessity, and our third contribution is a small but useful utility for Isabelle that allows the sledgehammer to be used in a completely unsupervised mode: the utility invokes sledgehammer on all the \lstinline{sledgehammer} commands in a given theory file, and if a proof is found, substitutes it into the theory file directly. 
We further improve on this by automatically invoking multiple \lstinline{sledgehammer} instances on a generated Isar (structured Isabelle) proof skeleton, and filling in the skeleton with the found proofs. We found this utility indispensable for completing our proof, and we have made it freely available for other Isabelle users working with similarly cumbersome proofs.

\paragraph{Auxiliary material}
Our Isabelle theory files, containing the definitions of our CXL model and the proof that it meets the SWMR property, are available on GitHub~\cite{ASPLOSGitHubRepo}.

\paragraph{Paper outline}
To provide intuition we first provide an overview of \CXLcache{} (\Cref{sec:overview}).
We then present salient details of our Isabelle formal model using standard mathematical notation (\Cref{sec:model}),
and describe the problems with the \CXLcache{} standard identified during the construction of this model, and our proposed fixes (\Cref{sec:fixingproblems}).
We then explain how we validated our model using scenario verification (\Cref{sec:scenarios}) and by proving an important coherence-related property---SWMR (\Cref{sec:proof}).
This large proof required some innovations in proof engineering, which we describe (\Cref{sec:automation}). We recap the assumptions and limitations on which our modelling and proof work is based (\Cref{sec:threats}), and discuss related and future work (\Cref{sec:related_future}).

\section{Overview of \CXLcache{}}
\label{sec:overview}

Before describing our formal model in detail (\Cref{sec:model}), we provide an intuitive overview of the \CXLcache{} protocol.

Multicore processors employ cache coherence protocols to ensure that multiple copies of the same data across different cores' caches remain in sync. With the rise of heterogeneous computing---where CPUs, GPUs, and specialized accelerators must work closely together---there is also a need for a global cache coherence protocol that manages data consistency across heterogeneous processors. This is the problem that \CXLcache{} is designed to solve.

\CXLcache{} allows devices like standalone GPUs and ASICs to cache a CPU's memory as if they are cores within the CPU's own multicore system.
 This facilitates correct and fine-grained data sharing with low latency, as the complex mechanisms that achieve cache coherence are managed at the hardware level.
For example, it might be desirable for an Intel CPU to be connected with an AMD GPU and an NVIDIA SmartNIC via some fast interconnect, allowing these accelerators to cache and share the CPU's memory within the same cache-coherent domain. If these devices have \CXLcache{} enabled, then this would be seamless. Capabilities like these are valuable for data centers that require composable infrastructure, where resources can be dynamically allocated and combined to optimize performance for diverse workloads. 

\definecolor{colorhost}{HTML}{A6CEE3}
\definecolor{colordev}{HTML}{FCCDE5}

\begin{figure}
\centering

\begin{tikzpicture}[
arr/.style = {->, line width=1pt},
]

\draw[draw=none, rounded corners, fill=colorhost](-0.7,-2.3) rectangle (0.7,2.7);
\draw[draw=none, rounded corners, fill=colordev](-3.8,-2.3) rectangle (-2.3,2.7);
\draw[draw=none, rounded corners, fill=colordev](3.8,-2.3) rectangle (2.3,2.7);
\node[anchor=south, text=white] at (0,-2.3) {host};
\node[anchor=south, text=white] at (-3.05,-2.3) {device 1};
\node[anchor=south, text=white] at (3.05,-2.3) {device 2};

\matrix[
matrix of math nodes,
column sep = 3mm,
row sep = 1mm,
nodes = {fill=white, draw=black, line width=1pt, font=\vphantom{bg}\scriptsize, inner sep=1mm,},
](m){
           & \HTDReqI  &         & \HTDReqII  &             \\
\DBufferI  & \HTDRspI  &         & \HTDRspII  & \DBufferII  \\
           & \HTDDataI &         & \HTDDataII &             \\         
\devcacheI &           & \hcache &            & \devcacheII \\
           & \DTHReqI  &         & \DTHReqII  &             \\
           & \DTHRspI  &         & \DTHRspII  &             \\
\DProgI    & \DTHDataI &         & \DTHDataII & \DProgII    \\[5mm]
           & \Counter \\
};
\draw[arr] (m-1-2) to[bend right] (m-2-1);
\draw[arr] (m-2-2) to (m-2-1);
\draw[arr] (m-3-2) to[bend right] (m-4-1);
\draw[arr] (m-2-1) to (m-4-1);
\draw[arr] (m-7-1) to[bend left] (m-5-2);

\draw[arr] (m-4-1) to[bend right] (m-5-2);
\draw[arr] (m-4-1) to[bend right] (m-6-2);
\draw[arr] (m-4-1) to[bend right] (m-7-2);

\draw[arr] (m-5-2) to[bend right] (m-4-3);
\draw[arr] (m-6-2) to[bend right] (m-4-3);
\draw[arr] (m-7-2) to[bend right] (m-4-3);

\draw[arr] (m-4-3) to[bend right] (m-1-2);
\draw[arr] (m-4-3) to[bend right] (m-2-2);
\draw[arr] (m-4-3) to[bend right] (m-3-2);

\draw[arr] (m-1-4) to[bend left] (m-2-5);
\draw[arr] (m-2-4) to (m-2-5);
\draw[arr] (m-3-4) to[bend left] (m-4-5);
\draw[arr] (m-2-5) to (m-4-5);
\draw[arr] (m-7-5) to[bend right] (m-5-4);

\draw[arr] (m-4-5) to[bend left] (m-5-4);
\draw[arr] (m-4-5) to[bend left] (m-6-4);
\draw[arr] (m-4-5) to[bend left] (m-7-4);

\draw[arr] (m-5-4) to[bend left] (m-4-3);
\draw[arr] (m-6-4) to[bend left] (m-4-3);
\draw[arr] (m-7-4) to[bend left] (m-4-3);

\draw[arr] (m-4-3) to[bend left] (m-1-4);
\draw[arr] (m-4-3) to[bend left] (m-2-4);
\draw[arr] (m-4-3) to[bend left] (m-3-4);
\end{tikzpicture}

\caption{An overview of our model of a two-device CXL state}
\label{fig:state_overview}
\Description{A block diagram showing our model of a two-device CXL state.}
\end{figure}

\CXLcache{} is an asymmetric protocol, with coherence-related messages between \emph{devices} (typically accelerators) all going via a central \emph{host} (CPU). \Cref{fig:state_overview} shows the main components of a CXL system with two devices, with arrows indicating the direction of the messages that are passed between them, and coloured backgrounds to indicate which components belong to a device and which belong to the host.

\CXLcache{} assumes that each device maintains the coherence of its own internal cache hierarchy, and hence is able to treat each device as having a single cache (\devcacheI{} and \devcacheII{}). The host also has a cache (\hcache). Each cacheline can be in one of four `stable' states: modified (write-access and dirty), exclusive (write-access and clean), shared (read-access), and invalid.

There are various channels from the host to a device (\textsf{H2D}) and from a device to the host (\textsf{D2H}) along which requests, responses and data can be sent. These channels are separated to allow them to be implemented with different latencies.
Transactions along these channels can be categorised as follows: 

\begin{itemize}

\item{\bf D2H Request:} A device may send the host a request for read-access (\RdShared) or write-access (\RdOwn) to a location.
    
\item{\bf H2D Request:} The host may need to invalidate the cacheline on a second device via a snoop-invalidate ($\SnpInv$) request.

\item{\bf D2H Response:} The second device may respond with $\RspIHitSE$ to report that it is invalidating its cacheline having previously enjoyed shared or exclusive access.

\item{\bf H2D Response:} Finally, the host replies to the first device with a $\GOShared$ or $\GOModified$ message to grant the desired access.

\end{itemize}

Our model of the CXL state in \Cref{fig:state_overview} also includes buffers, programs, and a counter, all of which will be explained in \Cref{sec:formalmodel:details}.

To improve performance and reduce latency, \CXLcache{} permits weaker ordering guarantees than traditional interconnects like PCIe \cite{pcie}. Specifically, it does not enforce ordering between different memory locations and provides minimal ordering on the same cache block \cite{tutorialDebendra}. 
This allows scenarios where a device reads updated data before a synchronization flag is set, enabling more concurrency on the same physical network.

\section{A formal model of CXL}
\label{sec:model}

Our formal model of CXL is expressed in the language of the Isabelle proof assistant~\cite{isabelle}, and is provided as a set of Isabelle theory files in our GitHub repository~\cite{ASPLOSGitHubRepo}.

In this section, we present salient details of the model (using mathematical notation rather than Isabelle syntax).
The model has been carefully constructed from our reading of the \CXLcache{} specification, and refined based on discussions with cache coherence experts from the CXL consortium.
The creation of the model led to us identifying and proposing fixes for several problems in \CXLcache{}, of which we elaborate on a selection in \Cref{sec:fixingproblems}.
We have validated our model using scenario verification (\Cref{sec:scenarios}) and mechanised proof (\Cref{sec:proof}).

\subsection{An overview of our CXL state model}\label{sec:formalmodel:overview}
The main components of our model are the host, the devices, and the channels between them, as already shown in \Cref{fig:state_overview}. 
The caches and channels in that figure are all directly taken from the CXL specification \cxlcitesection{3.2.1}. The program components (\DProgI{} and \DProgII{}) are an invention of ours---they are solely used to control the sequence of state transitions when exploring specific scenarios in \Cref{sec:scenarios}. They only serve to trigger coherence transactions, and do not modify locations or read out values. 
The standard does not specify how devices come up with unique transaction identifiers, 
so we use a simple, globally accessible counter ($\Counter$). The buffers (\DBufferI{} and \DBufferII{}) are another invention of ours; they are used to simulate the dependence between the H2D Response and H2D Request channels that is implied by the standard \cxlcitesection{3.2.5}. 

In an effort to keep the proof tractable, we have fixed the number of devices to two. This means that our proof cannot guarantee the absence of coherence violations that only manifest when three or more devices interact;
but for analyzing and prototyping purposes it is common to start with two devices~\cite{survey}.

\subsection{Details of our CXL state model}\label{sec:formalmodel:details}

\begin{figure}
\begin{math}
\begin{array}{ @{} r @{~} c @{~} l @{} }
\StableState & \eqdef & \{\EM, \SH, \I \} \\
\DeviceTransientState & \eqdef & \{\IMAD, \IMA, \IMD, \SMAD, \SMD, \SMA, \\
 & & \quad \ISD, \ISAD, \ISA,\MIA, \SIA, \IIA, \SIAC\}\\
\HostTransientState & \eqdef & \{\MAD, \MA, \MD, \SAD, \SD, \SA, \ID, \IB,\SB, \\
 & & \quad \MB \}\\
 \DeviceState &\eqdef& \DeviceTransientState \cup \StableState \\
\HostState &\eqdef& \HostTransientState \cup \StableState \\
 \HostCache & \eqdef & \lrecord \val : \Val, \State : \HostState \rrecord \\
 \DevCache & \eqdef & \lrecord \val : \Val, \State : \DeviceState \rrecord\\
\UTID & \eqdef & \mathbb{N} \\
 \DTHReqType & \eqdef & \{\RdShared, \RdOwn, \CleanEvict, \\
 & & \quad \DirtyEvict, \CleanEvictNoData\}\\
 \DTHReq & \eqdef &  \DTHReqType  \times \UTID \\ 
 \DTHRespType & \eqdef & \{\RspIHitSE, \RspIFwdM, \RspSFwdM \}\\
 \DTHResp & \eqdef &  \DTHRespType \times \UTID \\ 
 \HTDReqType & \eqdef & \{\SnpData, \SnpInv\}\\
 \HTDReq & \eqdef &  \HTDReqType \times \UTID \\ 
 \HTDRespType & \eqdef & \{\GO, \GOWritePull, \\
 & & \quad \GOWritePullDrop \}\\
 \HTDResp & \eqdef &  \HTDRespType \times \DeviceState \times \UTID \\ 
  \Data & \eqdef &   \UTID \times  \Val \\ 
    \Buffer & \eqdef & \HTDResp \cup \HTDReq \cup \{\bot\} \\
    \Instruction & \eqdef & \{\Load, \Store, \Evict\} \\
    \SystemState & \eqdef & \lrecord
   \begin{array}[t]{@{}l@{~}l@{}}
   \DProgI :& \List{\Instruction}, \\
   \DProgII :& \List{\Instruction}, \\
   \devcacheI :& \DevCache, \\
   \devcacheII :& \DevCache, \\
   \DTHReqI :& \List{\DTHReq}, \\
   \DTHReqII :& \List{\DTHReq}, \\
   \DTHRspI :& \List{\DTHResp}, \\
   \DTHRspII :& \List{\DTHResp}, \\
   \DTHDataI :& \List{\Data}, \\
   \DTHDataII :& \List{\Data}, \\
   \HTDReqI :& \List{\HTDReq}, \\
   \HTDReqII :& \List{\HTDReq}, \\
   \HTDRspI :& \List{\HTDResp}, \\
   \HTDRspII :& \List{\HTDResp}, \\
   \HTDDataI :& \List{\Data}, \\
   \HTDDataII :& \List{\Data}, \\
   \DBufferI :& \Buffer, \\
   \DBufferII :& \Buffer, \\
   \DProgI : & \List{\Instruction}, \\
   \DProgII : & \List{\Instruction}, \\   
   \hcache :& \HostCache, \\
   \Counter :& \mathbb{N} ~ \rrecord
   \end{array}


\end{array}
\end{math}

\caption{Our model of a CXL state}
\label{fig:states_defTan}
\Description{Our model of a CXL state}
\end{figure}

\Cref{fig:states_defTan} presents the type of each of the twenty components that appear in \Cref{fig:state_overview}. We now explain those types in detail.

There are three `stable' cacheline states. We do not track the distinction between exclusive state and modified state, because transitions between these states have no effect on ownership, and the 
SWMR property that we are interested in proving is phrased only in terms of ownership. So, we use $\EM$ when the cacheline is in either of these states, alongside $\Shared$ for shared and $\Invalid$ for invalid.

While a transaction is being carried out, a cacheline can be in one of several additional `transient' states, depending on whether it is held on a device ($\DeviceTransientState$) or on the host ($\HostTransientState$). For instance, $\IMAD$ refers to a cacheline that is awaiting an acknowledgement ($\STATE{A}$) and some data ($\STATE{D}$) in order to complete its transition from the invalid state ($\I$) to the modified state ($\EM$). 
These transient states are not officially part of the \CXLcache{} specification, so we follow the standard notation for them~\cite{primer}. 

A cacheline ($\HostCache$ or $\DevCache$) consists of a value ($\Val$) together with a stable or transient state. We are concerned with coherence, which is a property of a single memory location, so we assume without loss of generality that our caches contain just a single location.

Messages can be grouped into transactions, and each transaction has an identifier $\UTID$. A single transaction may involve several request and response messages; for instance, a device may send a request to the host, which requires the host to send a request to another device, which then responds to the host, finally allowing the host to respond to the first device.

A device can request from the host ($\DTHReq$) read-only access ($\RdShared$) or write access ($\RdOwn$).
Additionally, it can relinquish access to a location that has not been written ($\CleanEvict$) or that has been written ($\DirtyEvict$).
The message $\CleanEvictNoData$ is the same as $\CleanEvict$, but the device is additionally signalling that it will refuse to provide the (clean) data and the host must not request it.

There are additional device-to-host requests that we exclude from our model: $\RdCurr$ simply checks the current data value and does not affect ownership (nor coherence); $\RdAny$'s functionality is already covered by $\RdOwn$ and $\RdShared$; $\RdOwnNoData$ is no different from $\RdOwn$ from the perspective of the SWMR property;
and although messages $\ItoMWr$, $\WrCur$, $\CLFlush$, $\WOWrInv$, $\WOWrInvF$ and $\WrInv$ are interesting from a memory-ordering point of view, they are not interesting for coherence.

A host can respond to a device ($\HTDResp$) by sending a `global observation' message ($\GO$).
This signifies that the host believes the device's request has now been seen by all relevant parties, and can now be considered complete~\cxlcitesection{3.2.2.1}. If the device has sent an evict request, the host can respond by instructing the device to send its data to the host ($\GOWritePull$), or to discard its data ($\GOWritePullDrop$) \cxlcitesection{3.2.4.2.14}.
In all cases, a host-to-device response includes the new $\DeviceState$ that the device's cacheline should enter. 
There are additional host-to-device responses that we exclude from our model: $\WritePull$ is only used in response to $\WrInv$ requests which, as mentioned above, we do not model, and $\FastGOWritePull$ and $\ExtCmp$ provide an advanced optimisation where a device can indicate that an update is partially observable ($\FastGOWritePull$) and then globally observable ($\ExtCmp$). We currently do not model non-ideal network conditions or error-handling and therefore leave out the $\GOErrWritePull$ message too.

A host-to-device request ($\HTDReq$) is a snoop, used to check (and change) the status of the device's cacheline~\cxlcitesection{3.2.4.4}. If the request is a $\SnpData$, the device must downgrade its cacheline state to either $\SH$ or $\I$, and if the request is a $\SnpInv$, the device must downgrade its cacheline state to $\I$. In both cases, the device must send its data to the host if it is dirty. There exists also a $\SnpCur$ request for checking a device's cacheline without changing it, but we omit this from our model because it does not affect coherence.

A device-to-host response ($\DTHResp$) can be a $\RspIHitSE$ (which means that the device has downgraded from $\SH$ or $\Exclusive$ to $\I$~\cxlcitesection{3.2.4.3.3}), a $\RspIFwdM$ (which means that the device has downgraded from $\EM$ to $\I$ and is also forwarding its dirty data~\cxlcitesection{3.2.4.3.6}), or a $\RspSFwdM$ (which means that the device has downgraded from $\EM$ to $\SH$ and is also forwarding its dirty data~\cxlcitesection{3.2.4.3.5}). 
There are additional device-to-host responses that we exclude from our model. $\RspIHitI$ is not used because our model's host tracks device states and does not send out snoops unnecessarily. The transaction flows of $\RspVHitV$, $\RspSHitSE$
and $\RspVFwdV$ are very similar to those of $\RspSHitSE$, $\RspIHitSE$ and $\RspSFwdM$, respectively; we leave them out to avoid duplication in our proof.

Finally, each device's buffer ($\Buffer$) contains a single request or response message from the host, or is empty ($\bot$).


\subsection{CXL transitions}\label{sec:formalmodel:transitions}

\newcommand\rulename[1]{\textsc{#1}}

\newcommand\infrule[3]{
\begin{shaded*}
\noindent\raggedright
$\begin{array}{@{}l@{~}l@{}}
\multicolumn{2}{@{}l@{}}{\rulename{#1}} \\
\textbf{guards:}
#2
\textbf{actions:}
#3
\end{array}$
\end{shaded*}
}

\newcommand\infrulewarning[3]{
\begin{shaded*}
\noindent\raggedright
$\begin{array}{@{}l@{~}l@{}}
\multicolumn{2}{@{}l@{}}{\rulename{#1} $(\faWarning)$} \\
\textbf{guards:}
#2
\textbf{actions:}
#3
\end{array}$
\end{shaded*}
}

Our model consists of \rulecount{} rules that describe transitions between CXL states. \Cref{fig:infrules} presents a selection of these rules. Each rule consists of a name, a set of \textbf{guards} that must all hold in order for a rule to fire, and a set of \textbf{actions} by which some components of the state are (atomically) updated.

The \rulename{InvalidLoad1} rule says that if device 1's cache is in the invalid ($\Invalid$) state (first guard) and it wishes to perform a load (second guard), then it can request an upgrade to the shared ($\Shared$) state (first action), enter the $\ISAD$ state in the meantime (second action), and increment the transaction-identifier counter (third action).

\begin{figure}

\infrule{InvalidLoad1}{
& \devcacheI.\State = \Invalid \\
& \head(\DProgI) = \Load \\
}{
& \DTHReqI := \DTHReqI @ [(\RdShared, \\
& \quad\quad\quad\quad\quad\quad\quad\quad\quad\quad\quad\Counter )]\\
& \devcacheI.\State := \ISAD \\
& \Counter := \Counter + 1
}

\infrule{ModifiedStore1}{
& \devcacheI.\State = \Modified  \\
& \head(\DProgI) = \Store \\
}{
& \devcacheI.\Val := v\\
& \DProgI := \tail(\DProgI) \\
& \DBufferI := \EBuffer \\
& \Counter := \Counter + 1
}

\infrule{SharedSnpInv1}{
& \devcacheI.\State = \Shared  \\
& \head(\HTDReqI) =  (\SnpInv, \utid) \\
& \HTDRspI = [] \\
}{
& \devcacheI.\State := \Invalid\\
& \HTDReqI := \tail(\HTDReqI) \\
& \DBufferI := (\SnpInv, \utid) \\
& \DTHRspI := \DTHRspI @ [(\RspIHitSE, \utid)]
}

\infrule{HostModifiedDirtyEvict1}{
& \hcache.\State = \Modified  \\
& \devcacheI.\State = \MIA \\
& \head(\DTHReqI) = (\DirtyEvict, \utid) \\
& \HTDDataI = \DTHRspI = [] \\
}{
& \hcache.\State := \ID\\
& \DTHReqI := \tail(\DTHReqI) \\
& \HTDRspI := \HTDRspI @ [(\GOWritePull, \Invalid, \\
& \quad\quad\quad\quad\quad\quad\quad\quad\quad\quad\quad\utid)] \\
& \DBufferI := \EBuffer
}

\caption{A selection of our transition rules.}
\Description{A selection of our transition rules.}
\label{fig:infrules}
\end{figure}

The \rulename{ModifiedStore1} rule says that if device 1's cache is already in the modified ($\Modified$) state (first guard) and it intends to do a store (second guard), then no coherence messages are necessary; it need only write to its own cache (first action) and consider the instruction complete (second to fourth actions). To provide the reader with an intuitive store semantics we include the value written here, but we drop this during our proof because the SWMR property is independent of values; it cares only about ownership.

The \rulename{SharedSnpInv1} rule describes how a device deals with snoop requests from the host. If device 1's cache is in shared ($\Shared$) state (first guard) and the head of its H2D Requests channel is a $\SnpInv$ (guard 2), then its cache is invalidated (action 1), the H2D Request is removed (action 2) and put into the device's buffer (action 3), and a response is sent back to the host using the same transaction-identifier (action 4). This rule only fires if there are no outstanding H2D Responses (guard 3); this requirement captures the `Snoop-pushes-GO' rule, which dictates that an H2D Request (snoop) message cannot overtake an H2D Response (GO) message to the same device:
\begin{Quote}
    When the host returns a GO response to a device, the expectation is that a snoop
arriving to the same address of the request receiving the GO would see the results of
that GO.\hfill\cxlcitesection{3.2.5.2}
\end{Quote}\label{3252SPG}
\noindent We will show how relaxing this rule leads to a coherence violation in \Cref{sec:scenarios}.

\newcommand\hfilll{\hspace*{\fill}}

The \rulename{HostModifiedDirtyEvict1} rule describes a device requesting to evict dirty data. The rule fires if the host's cache is in modified ($\Modified$) state (first guard), the device's cache is in the process of changing state from modified ($\Modified$) state to invalid ($\Invalid$) state ($\MIA$, second guard), and the device has sent a $\DirtyEvict$ request (third guard). The host's cache enters the $\ID$ state because it will enter the invalid state once data arrives (first action), the D2H Request is removed (second action) and a corresponding H2D Response is issued (third action). 
The requirement that there are no H2D Data or D2H Response messages in-flight (fourth guard) is derived from the following `GO-cannot-tailgate-snoop' rule:
\begin{Quote}%
When the host is sending a snoop to the device, the requirement is that no GO response will be sent to any requests with that address in the device until after the Host has received a response for the snoop and all implicit writeback (IWB) data [\dots] has been received.\hfilll\cxlcitesection{3.2.5.2}
\end{Quote}\label{3252GTS}
\noindent
This requires the H2D Request, D2H Response, and D2H Data channels to contain no messages to the same address when sending a GO message.

\section{Fixing problems in the standard}\label{sec:fixingproblems}

The weight of industrial support behind CXL makes it likely that the standard will be implemented by multiple vendors over the coming years.
To ensure compatibility between implementations from different vendors, it is thus essential that the standard is precise and unambiguous.

Unfortunately, we have found that the current \CXLcache{} standard~\cite{cxl31} suffers from numerous inaccuracies and ambiguities.
We give some examples, which were discovered during the process of creating the formal model described in \Cref{sec:model}.
We have proposed fixes to address these shortcomings, and have discussed them with members of the CXL consortium who lead drafting of the \CXLcache{} protocol.
As detailed below, in most cases our proposed fixes have been agreed and will be adopted in future versions of the standard.

\subsection{Ambiguity/inaccuracy regarding multiple snoops}\label{sec:fixingproblems:inaccuracy}

We believe that the following rule
\begin{Quote}%
The host is only allowed to have one snoop pending at a time per cacheline address per device.\hfilll\cxlcitesection{3.2.5.5}
\end{Quote}

\noindent
is ambiguous because `per' appears more than once. If a host has two snoops pending, must they be to different addresses \emph{and} different devices? Or must they be to different addresses \emph{or} different devices? In fact, neither of these interpretations is quite correct, because what the rule does not mention that it can be legal to have multiple pending snoops on the same cacheline, as long as they belong to the same transaction.

\paragraph{Proposed fix} We propose to amend the text as follows:

\begin{Quote}%
\textdeleted{The host is only allowed to have one snoop pending at a time per cacheline address per device.}
\textadded{At no time is the host allowed to have two or more snoops on the same cacheline address pending, unless they use the same transaction identifier and target different devices.}
\end{Quote}

\noindent
Our proposal has tentatively been agreed by the CXL consortium and we are working with them to fine-tune the wording.

\subsection{Redundant rule about multiple snoops}\label{sec:fixingproblems:redundancy}

There is some redundancy between rules about sending multiple snoops to the same address. Specifically, the following rule:

\begin{Quote}11. The Host must not send a second snoop request to an address until all responses, plus data if required, for the prior snoop are collected.\hfilll\cxlcitesection{3.2.5.14}
\end{Quote}

\noindent
repeats what is already specified earlier (we note that we have faithfully transcribed quotes from the specification, which inconsistently capitalises ``host''):

\begin{Quote}%
The host must wait until it has received both the snoop response and all IWB data (if any) before dispatching the next snoop to that address.\hfilll\cxlcitesection{3.2.5.5}
\end{Quote}

\paragraph{Proposed fix} To avoid confusion, we propose removing Rule 11 from \S3.2.5.14. Our proposal has been accepted by the CXL consortium and is due to be adopted.

\subsection{Clarification about WritePull responses}\label{sec:fixingproblems:unclarity}

The following rule:

\begin{Quote}%
Conversely, the host may not launch a WritePull for a write until it has received the snoop response (including data in case of RspFwd) for any snoops to the pending write's address. \hfilll\cxlcitesection{3.2.5.3}
\end{Quote}

\noindent
enjoys a subtle interaction with a rule in \S3.2.5.2 that restricts the launching of GO messages. Restricting GO messages is almost enough on its own; the only `gap' that the restriction on WritePull messages fills relates to $\WrInv$ requests, to which hosts can respond with a WritePull rather than a GO. 

\paragraph{Proposed fix.} To clarify this subtlety to the reader, we suggest the following amendment:

\begin{Quote}%
Conversely, the host may not launch a WritePull \textadded{(in response to a $\WrInv$)} for a write until it has received the snoop response (including data in case of RspFwd) for any snoops to the pending write's address.
\end{Quote}

\noindent Our proposal has tentatively been agreed by the CXL consortium and we are working with them to fine-tune the wording.

\subsection{Potential optimisation when evicting stale data}\label{sec:fixingproblems:inefficiency}

CXL requires that:
\begin{Quote}%
if a device Evict transaction has been issued [\dots] but has not yet processed its WritePull from the host, and a snoop hits the writeback, the device must [\dots] set the Bogus field in all the D2H data messages sent to the host. The intent is to communicate to the host that [\dots] the data from the Evict is potentially stale.~\hfilll\cxlcitesection{3.2.5.4}
\end{Quote}

In other words, if a device has requested to evict some data, and the host has determined via a snoop that this data is already stale, then the device should not send the data back to the host; it should instead mark its data messages as `bogus'.
An alternative in this situation could be for the host to send a WritePullDrop to the device rather than a WritePull, which instructs the device not to send any data messages at all.
This could offer an efficiency gain by avoiding some D2H data traffic.

\paragraph{Proposed fix.} In Table 3-23 (``D2H Request (Targeting Non Device-attached Memory) Supported H2D Responses''), add a `$\star$' in the ``\DirtyEvict{} / \GOWritePullDrop{}'' cell. The meaning of the `$\star$' shall be:
\begin{Quote}%
\textadded{if the Host has been able to determine that the device's data is stale, by means of a prior snoop, then the Host may issue a \GOWritePullDrop{} rather than a \GOWritePull{}.}
\end{Quote}
\noindent
This proposal remains under discussion with the CXL consortium, who are evaluating its backward-compatibility and whether it represents a meaningful opportunity for improving performance.

\subsection{Other clarifications}\label{sec:fixingproblems:minor}

We have also discussed some other minor clarifications to the specification with the CXL consortium, such as adding a note to the beginning of the Device-to-Host Requests section~\cxlcitesection{3.2.4.2} to clarify that certain requests are only legal when the device's cache is in a certain state (as this is currently not explained until about 20 pages later~\cxlcitesection{3.2.5.15}).

\section{Scenario verification}
\label{sec:scenarios}
\begin{figure}

\usetikzlibrary{arrows.meta, positioning, shapes.geometric, decorations.markings, intersections, calc}

\tikzset{
    redtrans/.style={
        ->,
        draw=red,
        text=red
    },
    bluetrans/.style={
        ->,
        draw=blue,
        text=blue
    },
    main node/.style={
        draw,
        align=center,
        fill=white 
    }
}

\begin{tikzpicture}[>=Stealth, 
                    node distance=1cm, 
                    font=\sffamily]

\node at (3,10.7) {$\devcacheI.\State$};
\node at (6,10.7) {$\hcache.\State$};
\node at (9,10.7) {$\devcacheII.\State$};

\draw (3,4) -- (3,10) node[main node, above] {I};
\draw (6,4) -- (6,10) node[main node, above] {I};
\draw (9,4) -- (9,10) node[main node, above] {I};

\draw[bluetrans, name path=LineRdOwn] (3,9.5) -- (6,9) node[midway, above, sloped] {RdOwn};

\draw[redtrans, name path=LineGOSData] (6,9.6) -- (9,5) node[pos=0.2, above, sloped] {GO-S+Data};

\draw[redtrans, name path=LineRdShared] (9,10) -- (6,9.7) node[midway, above, sloped] {RdShared};

\draw[bluetrans, name path=LineSnpInv] (6,8.8) -- (9,7.5) node[pos=0.7, below, sloped] {SnpInv};

\draw[bluetrans] (9,7.5) -- (6,6) node[pos=0.7, above, swap, sloped] {RspIHitI};

\draw[bluetrans] (6,5.9) -- (3,5) node[midway, below, sloped] {GO-M+Data};

\path [name intersections={of=LineSnpInv and LineGOSData, by=I}];

\node[align=left] (violation) at (4.5,7) {Violation\\ occurs here};

\draw[->, thick] (violation.east) to [bend right] (I);

\draw[redtrans, dashed, ->] ($(I) + (0, 0.1)$) -- ($(9, 7.5) + (0, 0.1)$) node[midway, above, sloped] {Correct flow};

\node[main node] at (9,4.7) {S};
\node[main node] at (3,4.5) {M};

\end{tikzpicture}

\caption{A message sequence chart from~\cite{webinar} demonstrating a coherence violation if the snoop-pushes-GO rule is relaxed}
\Description{A message sequence chart.}
\label{fig:snooppushesGOrelaxation}

\end{figure}

In \Cref{sec:proof} we turn our attention to using Isabelle to prove that our model of \CXLcache{} satisfies the SWMR property.
Before that, in this section, we describe the \emph{scenario verification} activities we undertook before embarking on this proof.
These serve as important smoke tests to confirm that our formal model of \CXLcache{} actually behaves as one would expect in a variety of scenarios; if this were not the case then our formal proof would be meaningless.
Additionally, scenario verification allows us to scrutinise various restrictions that the \CXLcache{} protocol imposes to assess whether they are really necessary---i.e.\ whether relaxing these restrictions can lead to coherence violations.
This represents an important use case for our model beyond being a vehicle for formal proofs: it has the potential to allow protocol designers to rigorously understand the implications of the protocol rules and the consequences of relaxing them.

All three of the scenarios described in this section are produced in a semi-automatic way by Isabelle using its \lstinline|value| command. We provide the programs that the two devices should run, and give Isabelle a bound on the path depth to explore. We say \emph{semi}-automatic because in some cases it is also necessary to manually prune the tree of possible paths by adding extra predicates, in order to guide Isabelle towards a solution that we already have in mind. Without this, the search space can be so large that Isabelle does not terminate within a reasonable time bound.

\subsection{Smoke testing via litmus tests}\label{sec:scenarioverification:smoketesting}

To smoke test our model, we created a series of litmus tests.
Each litmus test initialises the system in a state where the two devices are poised to issue a particular series of requests, and confirms that, regardless of how nondeterminism in the transition rules is resolved, the model ends up in an expected final state and that no coherence violations occur in this or any intermediate states.
We illustrate this by describing two such litmus tests.

\begin{itemize}
\item{\bf Litmus test: \texttt{clean\_evict\_test}.}
\Cref{fig:clean_evict_sequence} illustrates this litmus test, showing the sequence of transitions that our model makes starting from an initial state. Intuitively, this test confirms that an eviction from a clean cache ends successfully.
In the initial state, both devices are in the $\Shared$ state, with device 1 having multiple evictions as instructions.
The transitions show that the first $\Evict$ first triggers a $\CleanEvict$
in the $\DTHReqI$ channel, and causes
a downgrade to the $\SIA$ state. 
Then the request is processed by the host.
The host sends a $\GOWritePullDrop{}$ message, which marks the  overall state of 
all caches as $\Shared$ since another device also has a copy at that point.
Finally, the device receives the $\GOWritePullDrop$ message, 
and downgrades to $\Invalid$ state. The $\Evict$ instruction is removed 
from the instruction list ($\DProgI$).
Subsequent $\Evict$s have no effect on $\devcacheI$ because it is already invalid.

\begin{table*}
\caption{A transition sequence witnessing \texttt{clean\_evict\_test}, a clean eviction from device 1.}
\resizebox{\linewidth}{!}{
$
\begin{array}{l@{~}c@{~}c@{~}c@{\hspace*{-20pt}}c@{~}c@{~}c@{~}c}
\toprule
\text{transition rule} & \DProgI & \devcacheI & \DTHReqI  & \HTDRspI & \hcache & \devcacheII & \Counter \\
\midrule
\text{(initial state)} & [\Evict, \Evict] & (0, \Shared) & [] & []  & (0, \Shared) & (0, \Shared) & 0 \\
\rulename{SharedEvict1} & [\Evict, \Evict] & (0, \SIA) & [(\CleanEvict, 1)]  & []  & (0, \Shared) & (0, \Shared) & 1 \\
\rulename{Shared\_CleanEvict\_NotLastDrop1} & [\Evict, \Evict] & (0, \SIA) & []  & [(\GOWritePullDrop{}, 1)]  & (0, \Shared) & (0, \Shared) & 1 \\
\rulename{SIA\_GO\_WritePullDrop1} & [\Evict] & (0, \Invalid) & [] & [] & (0, \Shared)  & (0, \Shared) & 1 \\
\rulename{SIA\_GO\_WritePullDrop1} & [\Evict] & (0, \Invalid) & [] & [] & (0, \Shared)  & (0, \Shared) & 1 \\
\bottomrule
\end{array}$
}
\label{fig:clean_evict_sequence}
\end{table*}

\item{\bf Litmus test: \texttt{dirty\_evict\_test}.}
Similarly, \Cref{fig:dirty_evict_sequence} illustrates the \texttt{dirty\_evict\_test} litmus test.
This test involves a sequence with a writer issuing a $\DirtyEvict$, to which the host responds with a $\GOWritePull$. This triggers a writeback from the eviction device, which the host copies in, marking the completion of this operation.

\end{itemize}

\begin{table*}
\caption{A transition sequence witnessing \texttt{dirty\_evict\_test}, a writeback triggered by $\GOWritePull$.}
$
\begin{array}{l@{~}c@{~}c@{~}c@{\hspace*{-20pt}}c@{\hspace*{-15pt}}c@{~}c@{~}c@{~}c}
\toprule
\text{transition rule} & \DProgI & \devcacheI & \DTHReqI & \DTHRspI & \HTDDataI & \hcache & \devcacheII & \Counter\\
\midrule
\text{(initial state)} & [\Evict] & (1, \Modified) & [] & [] & [] & (0, \Modified) & (0, \Invalid) & 0 \\
\rulename{ModifiedEvict1} & [\Evict] & (0, \MIA) & [(\DirtyEvict, 1)] & [] & [] & (0, \Modified) & (0, \Invalid) & 1 \\
\rulename{HostModifiedDirtyEvict1} & [\Evict] & (0, \MIA) & [] & [(\GOWritePull, 1)] & [] & (0, \ID) & (0, \Invalid) & 1 \\
\rulename{MIA\GO\_WritePull1} & [] & (1, \Invalid) & [] & [] & [(\const{Data}, 1)] & (1, \ID) & (0, \Invalid) & 1 \\
\rulename{IDData1} & [] & (1, \Invalid) & [] & [] & [] & (1, \Invalid) & (0, \Invalid) & 1 \\
\bottomrule
\end{array}$
\label{fig:dirty_evict_sequence}
\end{table*}

Our GitHub repository~\cite{ASPLOSGitHubRepo} includes 8 litmus tests that cover scenarios such as a read and a write being issued concurrently by two devices, multiple reads, multiple writes and multiple evicts, and alternating reads, writes and evicts.
We have evaluated all intermediate states in the execution traces, ensuring that tests complete successfully, maintaining a coherent state throughout.

\subsection{Assessing the \CXLcache{} restrictions}\label{sec:scenarioverification:restrictions}

Recall from \Cref{sec:formalmodel:transitions} that \CXLcache{} imposes various restrictions on implementations, such as the `Snoop-pushes-GO' rule that we discussed in relation to the \rulename{SharedSnpInv1} transition rule (\Cref{fig:infrules}).
These restrictions are formalised as predicates on system states that appear in the guards of various transition rules.

Because such restrictions place constraints on implementations of \CXLcache{}, one would reasonably expect that each of these restrictions is \emph{necessary}---i.e.\ that removing a restriction would compromise the correctness of the protocol.
We show that scenario verification using our Isabelle model can confirm this: that if a particular restriction is relaxed, additional states become reachable, and coherence violations can be observed.
This helps to establish confidence that CXL is not imposing restrictions unnecessarily.

We illustrate this for one such restriction.

\paragraph{Restriction test: \texttt{snoop\_pushes\_go\_test}}

\Cref{fig:goaftersnooprelaxation} shows how a coherence violation can be reached if the rule from \cref{3252SPG} that $\SnpInv$ messages cannot overtake $\GO$ messages is relaxed. The violation recreates a scenario that was explained in the form of a message-sequence chart in a CXL webinar~\cite{webinar}.

In the initial state, both devices' cachelines are invalid, and program 1 has a pending write and program 2 a pending read. 
Both devices start requests, and device 2's $\RdShared$ gets processed first, causing the host to send a $(\GO,\Shared)$ message and the associated data. Before device 2 takes these two messages, device 1's $\RdOwn$ gets processed, causing the host to send a $\SnpInv$ to the device, invalidating its cacheline.

\infrulewarning{ISADSnpInv2}{
& \devcacheII.\State = \ISAD  \\
& \head(\HTDReqII) =  (\SnpInv, \utid) \\
& \msout{\HTDRspII = []} \\
}{
& \HTDReqII := \tail(\HTDReqII) \\
& \DTHRspII := \DTHRspII @ [(\RspIHitI, \utid)] \\
& \DBufferII := (\SnpInv, \utid)
}

The modified \rulename{ISADSnpInv2} (\faWarning) rule above allows a snoop to be processed before the $\HTDRspII$ queue is empty.
The rest of the steps are just the host forwarding the response to device 1, and both devices taking the $\GO$ and data messages.
Observe that in the final row of \Cref{fig:goaftersnooprelaxation} both devices hold their cachelines in the $\Modified$ state, violating coherence.
In our correct model, $\devcacheII$ would not take the snoop until it has received the $\GO$ message and upgraded to $\ISD$.

\begin{table*}
\caption{A transition sequence witnessing \texttt{snoop\_pushes\_go\_test}, leading to an incoherent state if rule \rulename{ISADSnpInv2} is broken. In each row, $\DProgI = [\Store]$ and $\DProgII = [\Load]$.}
\resizebox{\linewidth}{!}{
$
\begin{array}{l@{~}c@{~}c@{\hspace*{-10pt}}c@{~}c@{~}c@{~}c@{\hspace*{-15pt}}c@{\hspace*{-12pt}}c@{~}c@{~}c@{~}c@{~}c@{~}c@{~}c}
\toprule
\text{transition rule} & \devcacheI & \DTHReqI & \HTDRspI & \HTDDataI & \hcache & \DTHReqII & \DTHRspII & \HTDReqII & \HTDRspII & \HTDDataII & \devcacheII & \Counter \\
\midrule
\text{(initial state)} &
(-1, \Invalid) & [] & [] & [] & (0,\Invalid) & [] & [] & [] & [] & [] & (-1, \Invalid) & 0 \\
\rulename{InvalidStore1} &
(-1, \IMAD) & [(\RdOwn, 0)] & [] & [] & (42,\Invalid) & [] & [] & [] & [] & [] & (-1, \Invalid) & 1 \\
\rulename{InvalidLoad2} &
(-1, \IMAD) & [(\RdOwn, 0)] & [] & [] & (42,\Invalid) & [(\RdShared, 1)] & [] & [] & [] & [] & (-1, \ISAD) & 2 \\
\rulename{InvalidRdShared2} &
(-1, \IMAD) & [(\RdOwn, 0)] & [] & [] & (42,\Shared) & [] & [] & [] & [(\GO, \Shared, 1)] & [(\Data(42),1)] & (-1, \ISAD) & 2 \\
\rulename{SharedRdOwn1} &
(-1, \IMAD) & [] & [] & [(\Data(42),0)] & (42,\MA) & [] & [] & [(\SnpInv, 0)] & [(\GO, \Shared, 1)] & [(\Data(42),1)] & (-1, \ISAD) & 2 \\
\rulename{ISADSnpInv2} (\faWarning) &
(-1, \IMAD) & [] & [] & [(\Data(42),0)] & (42,\MA) & [] & [(\RspIHitI, 0)] & [] & [(\GO, \Shared, 1)] & [(\Data(42),1)] & (-1, \ISAD) & 2 \\
\rulename{ISADGO+Data2} &
(-1, \IMAD) & [] & [] & [(\Data(42),0)] & (42, \MA) & [] & [(\RspIHitI, 0)] & [] & [] & [] & (42, \Shared) & 2 \\
\rulename{MARspIHitI1} &
(-1, \IMAD) & [] & [(\GO, \Modified, 0)] & [(\Data(42),0)] & (42, \Modified) & [] & [] & [] & [] & [] & (42, \Shared) & 2 \\
\rulename{IMADGO+Data1} &
(42, \Modified) & [] & [] & [] & (42, \Modified) & [] & [] & [] & [] & [] & (42, \Shared) & 2 \\ \bottomrule
\end{array}$
}
\label{fig:goaftersnooprelaxation}
\end{table*}

\section{Proving the SWMR property}
\label{sec:proof}

In this section, we present our proof that our model satisfies the Single-Writer-Multiple-Reader (SWMR) property.

The proof as a whole consists of \filecount{} theory files totalling around 211k lines of code. Most of these lines are taken up by \rulecount{} giant rule lemmas, each lemma  taking up about 2.5k lines of code with its \conjunctcount{} subgoals. It took us about 12 person-months to reach this. Most of the code has been generated by our \supersketch{} tool (as we explain in \Cref{sec:automation}), and only the definitions are purely handwritten, taking up less than ten thousand lines of code.
It takes approximately 1--2 minutes to check each rule file, and 3--5 hours to build a session consisting of all rule files on an Intel Core i9-14900HX running at 2.20 GHz.

The SWMR property states that if one device has write access ($\Modified$) to a location, then no other device can have read access ($\Shared$) or write access to the same location.

\begin{definition}[SWMR]
\begin{equation*}
 \wedge_{i \neq j}\, \neg\big( \dotfield{\devcachei{i}}{State} = \Modified \land \dotfield{\devcachei{j}}{State} \in  \{\Shared,\Modified\} \big)\ \label{eq:SWMR}
\end{equation*}
\end{definition}

\newcommand\evolvesto[1][]{\xrightarrow{#1}}
\newcommand\evolvestomany{\rightarrow^*}

\noindent Let us write $\Sigma \evolvesto \Sigma'$ if state $\Sigma$ can evolve in one step to state $\Sigma'$ via any of our transition rules.
Unfortunately SWMR is not inductive; that is, the following does not hold:
\[
    \text{If}~\Sigma \evolvesto \Sigma'~\text{and} ~\SWMR(\Sigma)~\text{then}~\SWMR(\Sigma').
\]
\noindent
A straightforward counterexample is a state that is about to become incoherent, such as:
\[
\llparenthesis \begin{array}[t]{@{~}l@{}}
\devcacheI = \llparenthesis 0, \IMA \rrparenthesis, \\
\HTDRspI = [\llparenthesis \GO, \Modified, \utid\rrparenthesis], \\
\devcacheII = \llparenthesis 0, \Modified \rrparenthesis
~\rrparenthesis
\end{array}
\]


\noindent 
However, 
this state is not reachable from any valid initial state.
We need a stronger property than SWMR to rule out erroneous and unreachable states like this one.
That is, we require an invariant $\inv$ such that:

\begin{itemize}
\item If $\texttt{initial\_state}(\Sigma)$ then $\inv(\Sigma)$.
\item If \smash{$\Sigma \makestep{}{} \Sigma'$} and $\inv(\Sigma)$ then $\inv(\Sigma')$.
\item If $\inv(\Sigma)$ then $\SWMR(\Sigma)$.
\end{itemize}

With this invariant in-hand, we can show that our system indeed satisfies the
SWMR property (writing $\evolvestomany$ for a sequence of zero or more transitions):

\begin{theorem}[SWMR\_CXL\_cache]
Assume that $\Sigma \evolvestomany \Sigma'$ and $\texttt{initial\_state}(\Sigma)$. Then $\SWMR(\Sigma')$.
\end{theorem}\label{thm:main}

The process of obtaining the invariant that enables this proof required a few dozen iterations to converge.
We started with SWMR and then successively added conjuncts to rule out erroneous and unreachable states as they became apparent. Whenever we added a conjunct, we sought to make it as simple and general as possible, in order to rule out as many bad states as possible in one go, while not excluding any reachable states.

We now present four of the conjuncts of $\inv$ to give the reader a flavour of the entire invariant. 

\paragraph{Transient states need similar SWMR constraints.}
The following conjunct of our invariant:
\[
\begin{aligned}
&\left(
    \begin{aligned}
    & \devcacheI.\State \in \{\IMD,\SMD\} \lor {}\\
    &  \devcacheI.\State \in \{ \IMAD,\, \SMAD \} \land \HTDRspI \neq []
    \end{aligned}
\right) \implies {}
\\[1ex]
&  \quad \head(\HTDReqII) \neq (\SnpInv,\_) \implies {} \\
&\left(
    \begin{aligned}
    & \devcacheII.\State \notin \{ \ISD,\, \IMD,\, \SMD,\, \ISA, \\
    & \quad\quad\quad\quad\quad\quad\quad\quad \IMA,\, \SMA,\, \Shared,\, \Modified \} \land {} \\
    & \HTDDataII = [] \land {} \\
    & \bigl( \devcacheII.\State \notin \{ \ISAD,\, \IMAD,\, \SMAD \} \lor{} \\
    & \quad \HTDRspII = [] \bigr)
    \end{aligned}
\right)
\end{aligned}
\]

\noindent says that if device 1 has almost upgraded to the $\Modified$ state, and is just awaiting an acknowledgement, then the other device must not be in any valid (or \emph{about to be} valid) states, unless a $\SnpInv$ is on its way to invalidate that valid cache.

\paragraph{Snoop responses need to be honest.}
If a device responds to a snoop that it has invalidated its cacheline, then it must, unsurprisingly, be in an invalid state:
\[
\begin{aligned}
& \head(\DTHRspI) \in \{(\RspIFwdM,\_), (\RspIHitSE,\_)\} \implies {} \\
& 
\devcacheI.\State \in \{\Invalid, \ISDI, \ISAD, \IMAD, \IIA\}
\end{aligned}
\]

\paragraph{Channels are singleton lists.}
As a result of our restriction to a single location, it is the case that each channel can contain at most one message at any given time:
\newcommand\length{\mathit{length}}
\begin{align*}
    &\length (\HTDReqI) \leq 1 \land
    \length(\HTDReqII) \leq 1 \land {} \\
    &\length (\HTDRspI) \leq 1 \land
    \length (\HTDRspII) \leq 1 \land {} \\
    &\length (\HTDDataI) \leq 1 \land
    \length (\HTDDataII) \leq 1 \land \dots 
\end{align*}

\paragraph{ Host and device data channels must not conflict. }

This is a stronger restriction than the previous conjunct. It says that each of the different data message channels $\htddatai{i}$ and $\dthdatai{j}$ has at most one data message pending:
\[ 
i \neq j\implies (\dthdatai{i} = [] \lor \htddatai{j} = []) 
\]

\section{Better automation for large proofs}
\label{sec:automation}

The difficulty associated working with large proofs of properties of computer systems is well known, and has been discussed e.g.\ in the context of the IronFleet project on proving correctness properties of distributed systems~\cite{IronFleet}, and the L4.verified project on verify an OS microkernel~\cite{ChallengesScalingLargeProofs}.
Proof scalability was a key challenge that we faced in working towards our proof of the SWMR property for our model of \CXLcache{}.
This is because deriving an inductive invariant that implied the SWMR property required many iterations of proof attempts, with each iteration taking a significant amount of human and machine time, and the time required increasing as the invariant grew.

We now outline the iterative process that we used to work towards an inductive invariant, explaining why this process was difficult and time-consuming (\Cref{sec:invariantdevelopment}).
We believe our report on this experience will be valuable for researchers interested in embarking on a formal verification project who do not yet have experience working on large inductive proofs.

We then describe a simple Isabelle utility, \supersketch{}, which we have created to accelerate the iterative development of inductive invariants (\Cref{sec:supersketch}).
This contribution is targeted more specifically at researchers intending to use the Isabelle prover for their verification efforts.

\subsection{The challenge of iterative inductive invariant development}\label{sec:invariantdevelopment}


Recall from \Cref{sec:proof} that our proof of the SWMR property hinges on an inductive invariant, $\inv{}$.
In practice, it was relatively easy to prove that the SWMR property was implied by $\inv$ and that all initial states of the system satisfied $\inv$.
Much more challenging was to prove that $\inv$ was actually inductive---i.e.\ that every successor of a state satisfying $\inv$ also satisfies $\inv$.

Viewing $\inv$ as a conjunction of sub-invariants, so that $\inv(\Sigma) =  \inv_1(\Sigma) \land \inv_2(\Sigma) \land \ldots \inv_n(\Sigma)$, we can treat the proofs we need to do to show the inductiveness of $\inv$ as an $n \times m$ matrix, where $n$ is the number of conjuncts
and $m$ is the number of transition rules.
Cell $(i, j)$ of this matrix represents the obligation to prove that $\inv(\Sigma) \implies \inv_i(\Sigma')$ whenever the transition $\Sigma \evolvesto \Sigma'$ is enabled by rule $j$ (we shall write $\Sigma \evolvesto[j] \Sigma'$ for this).
Demonstrating inductiveness involves generating proofs for all cells.
 
When we find that the proof for a cell $(i, j)$ does \emph{not} go through---i.e.\ we cannot prove that $\inv(\Sigma) \implies \inv_{i}(\Sigma')$ holds for $\Sigma \evolvesto[j] \Sigma'$---we \emph{strengthen} the invariant:
we devise a new conjunct $\inv_{n+1}$ such that we \emph{are} able to prove that $\inv(\Sigma) \land \inv_{n+1}(\Sigma) \implies \inv_{i}(\Sigma')$ holds for $\Sigma \evolvesto[j] \Sigma'$.

Let $\inv'$ denote the strengthened invariant we get by conjoining $\inv_{n+1}$ to $\inv$.
Having added this new conjunct means we must now prove that
$\inv'(\Sigma) \implies \inv_{n+1}(\Sigma')$ for all $\Sigma \evolvesto \Sigma'$, adding a new row consisting of cells $(n+1, 1), (n+1, 2), \ldots, (n+1, m)$ for rules $1$ to $m$ in our proof obligation matrix.
However, there is no guarantee that proofs for these new cells (other than for the $(n+1, j)$ cell) will go through.
If the proof for one such cell fails to go through we might add a new conjunct $\inv_{n+2}$ in response, introducing another row of proof obligations, which in turn may necessitate further conjuncts, and so on.

An even more problematic scenario is when we need to change or delete a conjunct $\inv_i$ from $\inv$ because it turned out to be incorrectly excluding valid states. In that case, we must invalidate not just the row $i$, but also any proofs that may depend on $\inv_i$.
Because we cannot be sure which proofs these are, it is necessary to re-check proofs for the entire matrix of proof obligations.

In practice, this iterative development proved to be very expensive both in terms of the machine time required to search for proofs and re-check existing proofs, and the human effort required when working with the proof assistant to coordinate this process.

\subsection{The \supersketch{} tool}\label{sec:supersketch}


When manually writing the rule lemmas for the obligation matrix discussed above, we observed that the proof obligations for most cells of the matrix were relatively simple to prove individually using Isabelle's automated proof-generation tool, \emph{sledgehammer}.
In our proof, we usually invoke dozens of sledgehammer calls simultaneously.
Sledgehammer is a rather expensive command: it encodes the current goal into solver inputs and invoke many instances of various solvers concurrently to maximize the chance of finding a proof quickly.
This means that one sledgehammer call can result in hundreds of automatic prover and SMT solver queries, and take several seconds to a minute to terminate.
Once a proof is found, a manual click is needed to adopt sledgehammer's proof into the script.
This needs to be done with care; for instance, we must adopt the proof from bottom to top, to prevent the insertion of earlier calls invalidating the results of later ones.

An easy automation step was to redirect sledgehammer's output to a file and then use a Python script to insert generated proofs into our overall proof script. 
Each lemma took 30-60 minutes to generate the proofs, but at least the process is automatic.
This allowed us to scale our invariant up to about 300 conjuncts. 

Still, we faced challenges when a simple sledgehammer call turned out to be insufficient for discharging a subgoal. 
In our proof we have around a dozen rules that require an initial case analysis to become tractable for Isabelle. In that case, we might need to split subgoals into sub-subgoals and make sledgehammer work on those. Making this automatic via external scripting at this nested depth of subgoal is rather clumsy and error-prone, and is very fragile under changes to the $\inv$ invariant. 

\newcommand\textsledge[1]{\textcolor{colorsledge}{\sethlcolor{colorsledgebg}\hl{#1}}}

\newcommand\Sketch{\texttt{Sketch}}

In response, we developed a tool, 
$\supersketch$, which breaks down a goal into (possibly) multiple subgoals using a method supplied by the user, concurrently calls sledgehammer on each of subgoal with several user-supplied heuristics, and finally generates a complete proof script with all the generated sub-proofs filled in.
This utility is based on Haftmann's \Sketch{} tool~\cite{sketch}. 
\Sketch{} generates a proof skeleton that shows what  needs to be proven for each subgoal (the \textcolor{colorsketch}{blue} text in \Cref{fig:supersketch}), leaving out the actual proofs for the user to manually put in.
What $\supersketch$ does in addition is that it invokes sledgehammer to search for proofs for the user (highlighted in \textsledge{pink} in \Cref{fig:supersketch}).

\begin{figure}
\begin{isabelle}
\isacommand{lemma} \\
\hspace*{1em}\isakeyword{assumes} "$\mathit{goal1} \land \mathit{goal2} \land \mathit{goal3} \land \mathit{goal4}$" \\
\hspace*{1em}\isakeyword{shows} "$\mathit{goal1'} \land \mathit{goal2'} \land \mathit{goal3'} \land \mathit{goal4'}$" \\
\isacommand{proof}\ - \\
\textcolor{colorsketch}{\isacommand{show}\ ?\isamath{thesis}} \\
\textcolor{colorsketch}{\hspace*{1em}\isacommand{proof}\ (intro\ conjI)} \\
\textcolor{colorsketch}{\hspace*{2em}\isacommand{show}\ $\mathit{g1p}$: "$\mathit{goal1'}$"} \textsledge{Proof 1 using smt queries} \\
\textcolor{colorsketch}{\hspace*{1em}\isacommand{next}} \\
\textcolor{colorsketch}{\hspace*{2em}\isacommand{show}\ $\mathit{g2p}$: "$\mathit{goal2'}$"} \textsledge{Proof 2 using metis} \\
\textcolor{colorsketch}{\hspace*{1em}\isacommand{next}} \\
\textcolor{colorsketch}{\hspace*{2em}\isacommand{show}\ $\mathit{g3p}$: "$\mathit{goal3'}$"} \textsledge{Proof 3 using smt queries} \\
\textcolor{colorsketch}{\hspace*{1em}\isacommand{next}} \\
\textcolor{colorsketch}{\hspace*{2em}\isacommand{show}\ $\mathit{g4p}$: "$\mathit{goal4'}$"} \textsledge{Proof 4 using smt queries} \\
\textcolor{colorsketch}{\hspace*{1em}\isacommand{qed}} \\
\isacommand{qed}
\end{isabelle}
\Description{An example proof.}
\caption{An example of our \supersketch{} utility}
\label{fig:supersketch}
\end{figure}

In the case where a subgoal cannot be solved automatically, $\supersketch$ emits a \lstinline{sorry} to indicate that no proof was found, in which case human intervention is required.
In our setting this happened less than 1\% of the time.
This tool allowed us to continue refining the inductive invariant from 300 conjuncts to almost 800, so that it finally converged. 

Although developed in response to our particular use case,
\supersketch{} can be applied more generally to prove Isabelle lemmas and theorems that can be efficiently broken down into subgoals that can be handled by automated theorem provers.
The idea on which \supersketch{} is based---closing the loop between a proof assistant (Isabelle in our setting) and a proof search tactic (sledgehammer in our setting)---could be applied in the context of other proof assistants.

We provide more details about the design and implementation of \supersketch{} in a separate paper~\cite{supersketch}.

\section{Assumptions and limitations}
\label{sec:threats}

We summarise the assumptions made by our work and corresponding limitations of our proof and modelling effort, most of which have been discussed earlier in the paper.

\paragraph{Restriction to coherence and the SWMR property}
Our proof efforts have focused on the SWMR property---a key part of proving coherence. We do not consider other properties such as deadlock freedom and liveness properties.
Because our focus is on coherence, we have not needed to model silent upgrades from the $\Exclusive$ to $\Modified$ state, and have hence collapsed these states together.

\paragraph{Two devices, one location}
As discussed in \Cref{sec:model}, our model considers two devices and one location.
Restricting to a single location is standard practice when reasoning about the SWMR property, which was our goal, but other properties such as deadlock-freedom and liveness would require consideration of multiple locations~\cite{primer}.
By restricting to two devices, certain scenarios are excluded such as invalidating multiple sharers and waiting for all their acknowledgements for an ownership-obtaining request.
A few rules and a fraction of the conjuncts in our inductive invariant rely on there being just two devices, e.g.\ if a device is upgrading to the $\Modified$ state, it can be immediately granted ownership if the other device's cache is in the $\Invalid$ state.
It should be straightforward to extend the model to cater for more devices, which would immediately allow more elaborate scenario verification (see \Cref{sec:scenarios}).
However, adapting our proof to this setting would require suitable abstraction and generalisation efforts.

\paragraph{A restricted set of \CXLcache{} messages}
As discussed in \Cref{sec:formalmodel:details}, our model omits certain device-to-host requests that form part of the \CXLcache{} protocol: $\RdCurr$, $\RdAny$, $\RdOwnNoData$, $\ItoMWr$, $\WrCur$, $\CLFlush$, $\WOWrInv$, 
$\const{WO}$
$\const{WrInvF}$, $\WrInv$, and $\CacheFlushed$~\cxlcitesection{3.2.4.2.5}.
We exclude these because they are not normally part of a cache-coherence protocol. $\RdCurr$ simply checks the most up-to-date data value and does not change cache state (and coherence). $\RdAny$'s functionality can be achieved by $\RdOwn$ and $\RdShared$ already. $\RdOwnNoData$ is no different to $\RdOwn$ from the perspective of the SWMR property.  $\WrCur$ does not affect coherence. $\ItoMWr$, $\CLFlush$, $\WOWrInv$, $\WOWrInvF$ and $\WrInv$ all require modelling more levels of memory hierarchies, and are left for future work.

\paragraph{Tracking mechanisms}
The host and devices sometimes need information about other system components to determine how to respond to messages.
For instance, in some cases the host needs to check whether the currently-evicting device is the last sharer in the system.
Such tracking is often achieved via bit-vectors in a physical implementation.
Our model assumes that the host does perfect tracking as if it can look at the state of the device caches (which would not be feasible in practice for efficiency reasons).
Currently \strengthenedrulecount{} rules rely on this ``perfect tracking'' assumption; details are in the \texttt{PerfectTrackingRules.txt} document in our GitHub repository~\cite{ASPLOSGitHubRepo}.

\paragraph{Additional validity threats}
We may have transcribed parts of the intention of the CXL standard incorrectly into Isabelle; however, we have mitigated this risk by carefully justifying where in the standard each of our modelling decisions comes from, and by consulting several experts on the CXL committee when we had doubts.
Finally, although a mechanised proof in a proof assistant such as Isabelle is considered the gold standard of correctness, it is always a remote possibility that software bugs in Isabelle or hardware bugs in the machine running it could undermine its guarantees.


\section{Related and future work}
\label{sec:related_future}

There have been several efforts to reason about a variety of cache-coherence protocols over the years~\cite{survey,arvind01}. Oswald et al. have developed a domain-specific language called ProtoGen~\cite{ProtoGen}, which automatically generates and verifies a cache-coherence protocol given a stable-state specification. It would be interesting to try to use their tool to specify \CXLcache{} and generate a complete protocol and then compare our model to that. Oswald et al. have also developed HeteroGen~\cite{HeteroGen}, which, like CXL, seeks to bridge heterogeneous systems for cache coherence, but uses ``proxy caches'' instead; it would be interesting to compare the functionalities of a CXL host and a proxy cache. Goens et al.~\cite{compound} have developed operational and axiomatic models for the memory ordering behind such heterogeneous systems. They abstract away the role of the interconnect; our work is complementary as we have modelled the interconnect in detail.

Hemiola~\cite{hemiola} is another domain-specific language for designing (and proving the correctness of) cache-coherence protocols over ordered networks. An important innovation is its `serializability proofs', whereby a user can prove properties about a system assuming transactions happen one by one, which can greatly reduce the number of concurrent situations to be considered. We believe that our proof could benefit from this technique once Hemiola has been extended to handle unordered networks; still, Hemiola would likely do little to reduce the complexity of our system-state invariant. However, Bourgeat hints at a way to reduce the complexity of invariants used for verifying pipelined processors~\cite[p.~131]{bourgeat_thesis} by making the invariant itself inductively-defined, and his approach may be adaptable to cache-coherence protocols.

In ongoing work, Assa, Friedman, and Lahav~\cite{assa+24} propose a model for programming on top of CXL. One of the assumptions they make is that CXL provides cache coherence. Our work can be seen as complementary to theirs in the sense that it helps to justify that assumption.

In future work, we would like to strengthen our theorem by relaxing our idealised tracking assumption. This will involve refining our inductive invariant further, and one mechanism for better managing the inevitable complexity of this would be to make the invariant more hierarchical---having more intermediate predicates between the top-level invariant and the atomic formulas. We would also like to extend our model to handle more than two devices, and to handle more than one location so that the memory \emph{consistency} model can be investigated~\cite{primer}. The sister protocol \CXLmem{}, which enables disaggregated memory~\cite{cxlmem}, is a natural target for future formalisation efforts too, and being a somewhat higher-level protocol, it should be amenable to more traditional litmus testing~\cite{litmus_tacas11}.

\section*{Acknowledgements}
We thank the anonymous reviewers for their valuable feedback, and Chris Hawblitzel for serving as our shepherd. This work was supported by an EPSRC Programme Grant on \emph{Interface reasoning for interacting systems} (EP/R006865/1). We thank Martin Desharnais and Jasmin Blanchette for valuable discussions, and for pointing us to the \Sketch{} tool, which inspired \supersketch{}. We thank Christian Urban for trying out \supersketch{} and providing useful feedback.
We thank Dan Iorga for sharing his insights on his investigation of CXL. We thank Vijay Nagarajan for sharing his expertise in cache coherence and heterogeneous protocols. We thank Debendra Sharma, Rob Blankenship and Thibaut Palfer-Sollier for helpful discussions related to the CXL standard.





\appendix
\section{Artifact Appendix}

\subsection{Abstract}

This artifact consists of the formal model of the cache coherence protocol of Compute Express Link (CXL)--\CXLcache{}. It contains the model and proof of the Single-Writer-Multiple-Reader (SWMR) property of \CXLcache{} in the Isabelle/HOL proof assistant, as described in the paper ``Formalising CXL Cache Coherence''.
The protocol is modelled as a transition system over system state, where a system state comprises cacheline states, communication channels, buffers and other auxiliary structures.

\subsection{Artifact check-list (meta-information)}


{\small
\begin{itemize}
  \item {\bf Algorithm:} \CXLcache{} Cache Coherence Protocol
  \item {\bf Program:} Isabelle theories
  \item {\bf Compilation:} isabelle jedit

  \item {\bf Run-time environment:} Windows, with Isabelle2023 installed. Double-click the ``Cygwin-Terminal'' .bat file in the installation folder, and run the command (in the Execution step) from that terminal.

    \item {\bf Execution:} First change into the artifact top-level directory. Then run the command ``isabelle jedit -l AllFixes -J -Xmx8192m'' in the artifact top-level directory.
  \item {\bf Output:} messages shown on the proof state panel indicating theory files have been successfully processed.

  \item {\bf How much disk space required (approximately)?:} 50GB
  \item {\bf How much time is needed to prepare workflow (approximately)?:} 10 minutes.
  \item {\bf How much time is needed to complete experiments (approximately)?:} 3-10 hours (It is recommended to use a machines with at least 32GB memory to achieve this lower bound.)
  \item {\bf Publicly available?:} Yes,

  \item {\bf Workflow automation framework used?:} Isabelle sessions.

\end{itemize}
}

\subsection{Description}

The file that contains the definitions of the system state with type-1 devices (corresponding to the datatype definition in \Cref{fig:states_defTan} in \Cref{sec:model}) is \transposed{} (see line 157 the record definition ``\typeonestate{}''). Together with the record type some functions for manipulating certain fields of \typeonestate{} are also defined.
The transition rules as shown in \Cref{fig:infrules} in \Cref{sec:model} of the system are defined in the file \buggyrules{}.

The coherence property that is shown to be an inductive invariant of the system lies in \coherenceproperties{} (see line 199, definition \swmrstatemachine{}).

The \basicinvariants{} file contains some basic invariants related to certain transitions and functions we already defined in \buggyrules{} and \transposed{}.

The proofs are in the rest of the .thy files in this artifact. Each transition rule is proven to maintain the inductive property (\swmrstatemachine{}). Since \swmrstatemachine{} is quite large (consisting of around 800 conjuncts), the proof of just a single rule is lengthy, each spanning more than 1,000 lines.
They are therefore each stored in a dedicated file, where the filename corresponds to the name of the rule (up to a prefix).

As an example, the \texttt{FixSIAGO\_WritePull.thy} file contains the proof that the \texttt{SIAGO\_WritePull} rule maintains the \swmrstatemachine{} property. The main lemma stating this fact is at the end of the file (line 2915 with name \texttt{SIAGO\_WritePull\_coherent}).

The most important auxiliary lemma leading to this is \texttt{SIAGO\_WritePull'\_coherent\_aux\_simpler} (see line 233). This auxiliary lemma breaks down the proof into hundreds of subgoals. We call lemmas like this ``rule lemmas'' as they each correspond to a rule.

The top level theorem stating the Single-Writer-Multiple-Reader property of the transition system is the corollary named \texttt{SWMR\_pplus\_cache} in \texttt{TopLevelTheorem.thy} (line 354). It corresponds to Theorem 6.2 in the paper.
Some main theorems lead to this corollary:

\begin{itemize}

\item If $\texttt{initial\_state}(\Sigma)$ then $\swmrstatemachine(\Sigma)$ \\ (Theorem \texttt{SWMR\_state\_machine\_CXL\_cache}, line 321).

\item If \smash{$\Sigma \makestep{}{} \Sigma'$} and $\swmrstatemachine(\Sigma)$ then \\ $\swmrstatemachine(\Sigma')$ \\ (Theorem \texttt{all\_transitions\_coherent}, line 103).
\end{itemize}

These two theorems correspond to the first two of the three properties described in the paper just before \Cref{thm:main}.

\subsubsection{How to access}

The artifact is available on GitHub:

\begin{small}
\url{https://github.com/ChengsongTan/CXLcacheFormalisation}
\end{small}

\subsubsection{Software dependencies}

The artifact depends on Isabelle2023, available at:

\begin{small}
\url{https://isabelle.in.tum.de/website-Isabelle2023/index.html}
\end{small}



\subsection{Installation}
See the ``Running experiment'' section on GitHub:

\noindent
\begin{small}
    \url{https://github.com/ChengsongTan/CXLcacheFormalisation?tab=readme-ov-file#running-experiment}
\end{small}
\subsection{Evaluation and expected results}
See the ``Expected results'' section on GitHub:

\noindent
\begin{small}
    \url{https://github.com/ChengsongTan/CXLcacheFormalisation?tab=readme-ov-file#expected-results}
\end{small}



    
    
    





\bibliographystyle{plain}
\balance
\bibliography{cxl.bib}

\end{document}